\documentclass[longauth, traditabstract, letter]{aa} 
\usepackage{graphicx}
\usepackage{txfonts}
\begin{document}
%
   \title{Herschel/HIFI observations of
interstellar OH$^+$ and H$_2$O$^+$ towards W49N\thanks{Herschel is an ESA space observatory with science instruments provided
by European-led Principal Investigator consortia and with important participation from NASA.}:
a probe of diffuse clouds with a small molecular fraction}

   \author{D. A. Neufeld\inst{1}, J.~R.~Goicoechea \inst{2}, P.~Sonnentrucker\inst{1}, J.~H.~Black \inst{3}, J.~Pearson \inst{4}, S.~Yu \inst{4}, T.~G.~Phillips \inst{5}, D.~C.~Lis \inst{5},  M.~De~Luca \inst{6},  E.~Herbst \inst{7}, P.~Rimmer\inst{7}, M.~Gerin \inst{6}, T.~A.~Bell\inst{5},  F.~Boulanger \inst{8}, J.~Cernicharo \inst{2}, A.~Coutens\inst{9}, E.~Dartois \inst{8}, M.~Kazmierczak\inst{10}, P.~Encrenaz \inst{6}, E.~Falgarone \inst{6}, T.~R.~Geballe \inst{11}, T.~Giesen\inst{12}, B.~Godard\inst{6},  P.~F.~Goldsmith \inst{4}, C.~Gry \inst{13}, H.~Gupta\inst{4}, P.~Hennebelle \inst{6}, P.~Hily-Blant\inst{14}, C.~Joblin\inst{9}, R.~Ko{\l}os \inst{15}, J.~Kre{\l}owski\inst{10}, J.~Mart\'in-Pintado\inst{2}, K.~M. Menten\inst{16}, R.~Monje\inst{5}, B.~Mookerjea \inst{17},   M.~Perault \inst{6}, C.~Persson \inst{3}, R.~Plume \inst{18}, M.~Salez \inst{6}, S.~Schlemmer\inst{12}, M.~Schmidt\inst{19}, J.~Stutzki \inst{12}, D.~Teyssier\inst{20}, C.~Vastel \inst{9},  A.~Cros \inst{9},  K.~Klein \inst{21}, A.~Lorenzani \inst{22}, S.~Philipp \inst{23}, L.~A.~Samoska \inst{4}, R.~Shipman \inst{24},
A.~G.~G.~M. Tielens \inst{25},  R.~Szczerba  \inst{19} \and J.~Zmuidzinas \inst{5}}     

   \institute{The Johns Hopkins University, Baltimore, MD 21218, USA \\ 
 \email{neufeld@pha.jhu.edu}
\and Centro de Astrobiolog\'{\i}a, CSIC-INTA, 28850, Madrid, Spain 
\and Chalmers University of Technology, G\"oteborg, Sweden 
\and JPL, California Institute of Technology, Pasadena, USA 
\and California Institute of Technology, Pasadena, CA 91125, USA 
\and LERMA, CNRS, Observatoire de Paris and ENS, France 
\and Depts.\ of Physics, Astronomy \& Chemistry, Ohio State Univ.\, USA 
\and Institut d'Astrophysique Spatiale (IAS), Orsay, France 
\and Universit\'e Toulouse; UPS ; CESR ; and CNRS ; UMR5187, 
9 avenue du colonel Roche, F-31028 Toulouse cedex 4, France 
\and Nicolaus Copernicus University, Torun, Poland 
\and Gemini telescope, Hilo, Hawaii, USA 
\and I. Physikalisches Institut, University of Cologne, Germany 
\and LAM, OAMP, Universit\'e Aix-Marseille \& CNRS, Marseille, France 
\and Laboratoire d'Astrophysique de Grenoble, France 
\and Institute of Physical Chemistry, PAS, Warsaw, Poland 
\and MPI f\"ur Radioastronomie, Bonn, Germany 
\and Tata Institute of Fundamental Research, Homi Bhabha Road, Mumbai 400005, India 
\and Dept. of Physics \& Astronomy, University of Calgary, Canada 
\and Nicolaus Copernicus Astronomical Center, Poland 
\and European Space Astronomy Centre, ESA, Madrid, Spain 
\and Department of Physics and Astronomy, University of Waterloo, Canada 
\and Osservatorio Astrofisico di Arcetri-INAF- Florence - Italy 
\and Deutsches Zentrum f\"ur Luft- und Raumfahrt e. V., Raumfahrt-Agentur,
Bonn, Germany 
\and SRON Netherlands Institute for Space Research, Netherlands 
\and Sterrewacht Leiden, Netherlands 
} 
  \abstract
{We report the detection of absorption by interstellar hydroxyl cations and water cations, along the sight-line to the bright continuum source W49N. We have used {\it Herschel}'s HIFI instrument, in dual beam switch mode, to observe the 972~GHz $N=1-0$ transition of $\rm OH^+$ and the 1115~GHz $1_{11}-0_{00}$ transition of ortho-$\rm H_2O^+$.  The resultant spectra show absorption by ortho-$\rm H_2O^+$, and strong absorption by OH$^+$, in foreground material at velocities in the range 0 to 70$\,\rm km\,s^{-1}$ with respect to the local standard of rest.  The inferred $\rm OH^+/H_2O^+$ abundance ratio ranges from $\sim 3$ to $\sim 15$, implying that the observed OH$^+$ arises in clouds of small molecular fraction, in the $2 - 8\%$ range.  
This conclusion is confirmed by the distribution of OH$^+$ and H$_2$O$^+$ in Doppler velocity space, which is similar to that of atomic hydrogen, as observed by means of 21 cm absorption measurements, and dissimilar from that typical of other molecular tracers.  The observed $\rm OH^+/H$ abundance ratio of a few $\times 10^{-8}$ suggests a cosmic ray ionization rate for atomic hydrogen of $0.6 - 2.4 \times 10^{-16}\rm s^{-1}$, in good agreement with estimates inferred previously for diffuse clouds in the Galactic disk from observations of interstellar H$_3^+$ and other species.}  

   \titlerunning{Interstellar OH$^+$ and $\rm H_2O^+$ along the sight-line to W49}
	\authorrunning{Neufeld, Goicoechea, Sonnentrucker et al.}
   \maketitle
%

\section{Introduction}

After hydrogen and helium, oxygen is the most abundant heavy element in the Universe, accounting for more than $1\%$ of the mass of baryons in the Galaxy.  Of the $\sim 150$ distinct molecules detected in the interstellar gas to date, more than one-quarter contain oxygen, including many of the most widely-observed species such as CO, OH, H$_2$O, CH$_3$OH, H$_2$CO, SiO, and SO$_2$.  Not surprisingly, the chemical processes leading to the incorporation of interstellar oxygen atoms into oxygen-bearing molecules have been the subject of numerous theoretical investigations.  These studies have identified three types of process that can be important in various interstellar environments: (1) in warm regions, where the gas temperature exceeds $\sim 300$~K as a result of heating by shocks, by turbulent dissipation, or by infrared or ultraviolet radiation, the oxygen chemistry can be initiated by the endothermic reaction of atomic oxygen with H$_2$; (2) on grain surfaces, the formation of oxygen-bearing molecules can occur following the adsorption of atomic oxygen; the resultant molecules can then be released into the gas-phase by photodesorption or evaporation if the grains are exposed to ultraviolet radiation or heated sufficiently; and (3) in gas clouds that are irradiated by cosmic rays or X-rays, the oxygen chemistry is controlled by a series of ion-neutral reactions, and is initiated by the formation of the hydroxyl cation OH$^+$, which can react with molecular hydrogen to form H$_2$O$^+$ and then H$_3$O$^+$.

This third type of process, ion-neutral chemistry, is believed to dominate the formation of oxygen-bearing molecules within the cold quiescent interstellar medium (ISM).  However, the detection of two key intermediaries in this reaction sequence, the OH$^+$ and H$_2$O$^+$ molecular ions, has proven elusive.  Like other light hydrides with small moments of inertia, these molecules have rotational transitions at high frequencies that are difficult or impossible to observe from ground-based observatories.  Until very recently, neither OH$^+$ nor H$_2$O$^+$ had been detected in the interstellar gas (although optical transitions of H$_2$O$^+$ have long been detected in comets, e.g.\ Wehinger et al.\ 1974). In the last three months, however, the observational picture has changed rapidly.  With the availability of the HIFI instrument (de Graauw et al.\ 2010) on {\it Herschel} (Pilbratt et al.\ 2010), detections of H$_2$O$^+$ have been reported in absorption toward DR21, Sgr B2 (M), NGC6334I (Ossenkopf et al.\ 2010) and G10.6 -- 0.4 (Gerin et al.\ 2010).  In G10.6--0.4, which was targeted for absorption line studies as part of the PRISMAS (``PRobing InterStellar Molecules with Absorption line Studies'') key program, absorption by $\rm H_3O^+$ and strong absorption by OH$^+$ were also reported.  In addition, {\it Herschel}/SPIRE observations of Mrk 231 have revealed luminous OH$^+$ emission (van der Werf et al.\ 2010), 
while recent ground-based observations have led to detection of interstellar OH$^+$ in the near-UV spectral region
(Kre{\l}owski, Beletsky \& Galazutdinov 2010; who used the ESO Paranal observatory to detect a weak absorption line at 358.3769 nm) and at submillimeter wavelengths (Wyrowski et al.\ 2010, who used the APEX observatory to detect OH$^+$ in absorption toward Sgr B2). 

In this {\it Letter}, we report {\it Herschel}/HIFI observations of OH$^+$ and  $\rm H_2$O$^+$  towards a second strong continuum source that we have targeted in the PRISMAS program, W49N.  This luminous region of star formation, lying within the W49A complex, is located at a distance of 11.4 kpc, a value reliably determined by means of maser proper motion studies (Gwinn et al.\ 1992).  The sight-line to W49N is known to intersect a large collection of foreground gas clouds, with velocities relative to the local standard of rest (LSR) ranging from $\sim 0$ to $\sim 70 \rm \, km\, s^{-1}$; these have been widely detected by means of absorption line spectroscopy of multiple species, including H (Brogan \& Troland 2001, Fish et al.\ 2003), H$_2$O, OH, (Plume et al.\ 2004), O (Vastel et al.\ 2000), HCO$^+$, HCN, HNC, and CN (Godard et al.\ 2010).


\begin{figure}
\includegraphics[width=9.2 cm]{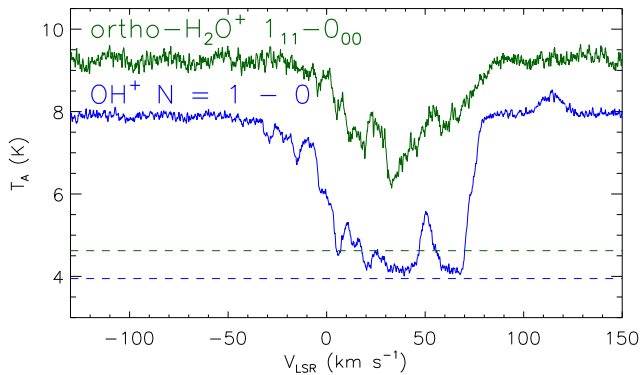}
\caption{Spectra of H$_2$O$^+$ $1_{11}-0_{00}$ ({\it green}) and OH$^+$ $N=1-0$ ({\it blue}) transitions obtained toward W49N.  The velocity scale applies to the strongest hyperfine component, with assumed frequencies of 971803.8~MHz for OH$^+$ and 1115204~MHz for $\rm H_2O^+$. Note that because HIFI employs double sideband receivers, the complete absorption of radiation at a single frequency will reduce the measured antenna temperature to one-half the apparent continuum level.}
\end{figure}
\section{Observations and data reduction}

In the observations reported here, we targeted two transitions -- the 972~GHz $N=1-0$ transition of OH$^+$, and the 1115~GHz $1_{11}-0_{00}$ transition of ortho-$\rm H_2O^+$  -- in the lower sidebands of Bands 4a and 5a of the HIFI receiver. 
To help determine whether any observed feature did indeed lie in the expected sideband, three observations were carried out with slightly different settings of the local oscillator (LO) frequency.  These observations, with on-source integration times of  59~s (OH$^+$ line) and 89~s ($\rm H_2O^+$ line) each, were carried out on 2010 April 18, using the dual beam switch (DBS) mode and the Wide Band Spectrometer (WBS).  The WBS has a spectral resolution of 1.1~MHz, corresponding to a velocity resolutions of $0.34\, \rm km\,s^{-1}$ to $0.30\, \rm km\,s^{-1}$ at the frequencies of the OH$^+$ and $\rm H_2O^+$ transitions. The telescope beam was centered at $\alpha =$19h10m13.2s, $\delta =$ 09$^{\circ}$06$^{\prime}$12.0$^{\prime\prime}$ (J2000).  
The reference positions for these observations were located 3$^\prime$ on either side of the source along an East-West axis.

The spectroscopic parameters for the observed transitions have been summarized by Ossenkopf et al.\ (2010) and Gerin et al.\ (2010).   {Both the OH$^+$ and H$_2$O$^+$ transitions exhibit hyperfine structure associated with the interaction of the nuclear spin magnetic moment with that resulting from electronic spin.}
The OH$^+$ $N=1-0$ transition has three hyperfine components, at velocities of --35.6, --0.5, and 0~km~s$^{-1}$ relative to the central hyperfine component, and with absorption line strengths in the ratio 1~:~5~:~9 if the lower levels are populated in proportion to their statistical weights.  For H$_2$O$^+$, there are five hyperfine components; the corresponding velocity shifts are $-25.5$, --15.9, 0, 4.8 and 14.4~km~s$^{-1}$ and the ratio of line strengths is 1~:~8~:~27~:~8:~10 .  Frequencies accurate to 1.5 MHz or better are available for the OH$^+$ transition, but those for the $\rm H_2O^+$ transition have been the subject of some controversy (Ossenkopf et al.\ 2010). As discussed in Appendix A, we favor a frequency close to that given by M\"urtz et al.\ (1998) for the strongest hyperfine component, $1115.204 \pm 0.002$~GHz.

The data were processed using the standard HIFI pipeline to Level 2, providing fully calibrated spectra of the source.  The Level 2 data were analysed further using the Herschel Interactive Processing Environment (HIPE; Ott 2010), version 2.4, along with ancillary IDL routines that we have developed.  For each of the target lines, the signals measured in the two orthogonal polarizations were in excellent agreement, as were spectra obtained at the three LO settings when assigned to the expected sideband.  We combined the data from the three observations, and from both polarizations, to obtain an average spectrum.

\section{Results}

Figure 1 shows the WBS spectra for the OH$^+$ $N=1-0$ (blue) and ortho-H$_2$O$^+$ $1_{11}-0_{00}$ (green) transitions, with the frequency scale expressed as Doppler velocities relative to the Local Standard of Rest (LSR) for the strongest hyperfine component. The double sideband continuum antenna temperatures are $T_A({\rm cont}) = 7.89$~K and 9.24~K respectively for the OH$^+$ and ortho-H$_2$O$^+$ transition, and the r.m.s noise values are 0.069~K and 0.117~K.  Because HIFI employs double sideband receivers, the complete absorption of radiation at a single frequency will reduce the measured antenna temperature to rougly one-half the apparent continuum level (given a sideband gain ratio $\sim$ unity).  Horizontal lines in Figure 1 indicate the values of $T_A({\rm cont})$ and $0.5\,T_A({\rm cont})$.  The OH$^+$ absorption feature shows a flat bottom in the LSR velocity intervals [35,42] and [60,70] km~s$^{-1}$ that does not {\it exactly} fall to $0.5\,T_A({\rm cont})$.  This behavior suggests that the sideband gain ratio (SBR) is not exactly unity, and we therefore treat the SBR as an adjustable parameter in the line fitting procedure described below.

In fitting the observed spectra, we have assumed the absorption to arise in a set of clouds with Gaussian opacity profiles.  Moreover, given the close chemical relationship between the two cations that we have detected, we assumed that the OH$^+$ and ortho-$\rm H_2O^+$ arise within the same set of clouds.  Each cloud is therefore characterized by a set of four quantities that we treat as adjustable parameters: the centroid velocity, v$_{LSR}$, the velocity width, $\Delta$v (defined here as the full-width-at-half-maximum), and the OH$^+$ and $\rm H_2O^+$ column densities.  Here, we neglected any contribution from para-H$_2$O$^+$ to the total $\rm H_2O^+$ column density; the $1_{01}-1_{01}$ transition of para-$\rm H_2O^+$ was searched for but not detected toward W49, and observations of para-$\rm H_2O^+$ toward Sgr B2 (M), reported recently by Schilke et al.\ (2010), imply that the ortho-to-para ratio is typically large ($\sim 5$) in diffuse clouds along that sight-line.

We also assumed that the population of OH$^+$ and ortho-$\rm H_2O^+$ in excited rotational states is negligible, and that the individual hyperfine states within the ground rotational state are populated in proportion to their statistical weights.  Our line fitting procedure included two additional free parameters: (1) the SBR for the OH$^+$ observations; and (2) the rest frequency of the H$_2$O$^+$ transition (see Appendix A)   We adopted the minimum number of absorption components needed to provide a reasonable fit to the data -- six -- and obtained initial estimates of the cloud velocities and FWHM's by eye.  These estimates were then refined by using the IDL mpfitfun routine, which implements the Levenberg-Marquardt method for obtaining a least-squares fit.  

Figure 2 shows the best fit that could be obtained to the OH$^+$ (top panel) and H$_2$O$^+$ (middle panel) spectra for a model with six foreground clouds.  The reduced $\chi^2$ for this fit was 1.83, suggesting that the ``minimal" absorption model does not entirely account for the data.  
The best-fit cloud parameters are given in Table 1.  {The best-fit line widths are moderately broad, particularly for the 62.5 km~s$^{-1}$ feature, and might -- in reality -- represent the superposition of multiple narrower features.} The SBR derived for the OH$^+$ spectrum was $0.884$ (signal sideband gain divided by image sideband gain), and the frequency derived for the strongest H$_2$O$^+$ hyperfine component was $1115209$ MHz (see Appendix A).   In the bottom panel of Figure 2, we present the OH$^+$ (blue) and H$_2$O$^+$ column densities (green) obtained per unit velocity width in our best fit model.  The cloud parameters in Table 1 appear to be relatively robust.  Although the OH$^+$ absorption is completely thick for LSR velocities in the ranges 30 -- 42 and 60 -- 70 km~s$^{-1}$ (as determined for the strongest hyperfine component), the existence of a weaker hyperfine component at an offset of --35 km~s$^{-1}$ means that arbitrarily large column densities of OH$^+$ can only be accommodated within a narrow range of LSR velocities: 65 -- 70  km~s$^{-1}$. 

\begin{table}
\caption{W49N absorption line components}
\begin{tabular}{c c c c c c c  }
\hline\hline
$v_{\rm LSR}$ & $\Delta v$ & $N({\rm OH^+})$ & $N({\rm H_2O^+})$ & $\rm OH^+/ H_2O^+$ & $f({\rm H_2})\,^a$ \\
$\rm km\,s^{-1}$ & $\rm km\,s^{-1}$ & $10^{13}\rm cm^{-2}$ & $10^{13}\rm cm^{-2}$ \\
\hline
6.6   &  5.1   &  2.9 & 0.42 &  6.8 & 0.034 \\
13.4  &  4.6   &  2.2 & 0.69 &  3.2 & 0.082 \\
21.2  &  7.0   &  8.2 & 0.56 &  14.6 & 0.015 \\
34.7  &  9.0   &  26.1 & 2.68 &  9.7 & 0.023 \\
43.5  &  8.4   &  7.5 & 0.72 &  10.5 & 0.021 \\
62.5  &  13.3  &  18.9 & 1.67 &  11.3 & 0.019 \\
\hline\hline

\end{tabular}
$^a$using eqn.\ (1) for an assumed temperature of 100~K and an assumed electron abundance of $1.4 \times 10^{-4}$.  The required molecular fraction, $f({\rm H}_2)$, scales as $T^{-0.5}\,x_e$
\end{table}

\begin{figure}
\includegraphics[width=9cm]{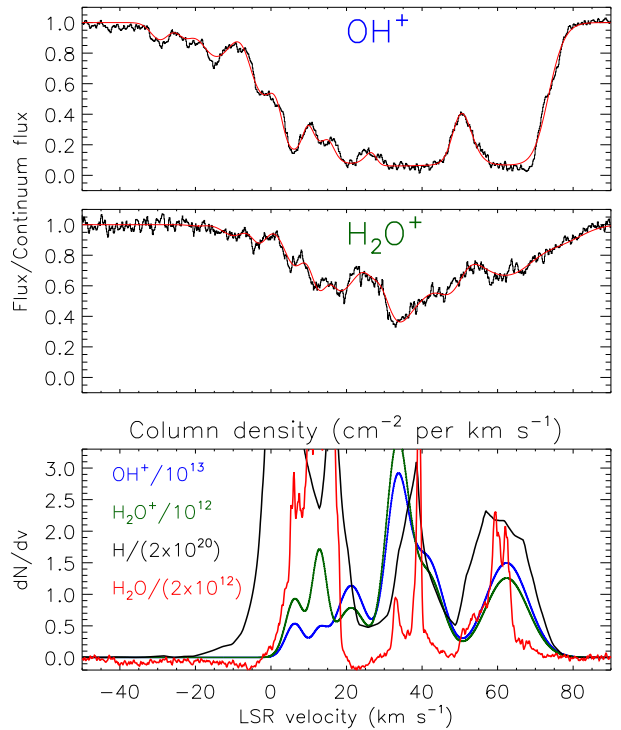}
\caption{Fit to the OH$^+$ ({\it top panel}) and H$_2$O$^+$ ({\it middle panel}) absorption spectra observed toward W49N.  The black curve shows the data, and the red curve is the fit obtained for the cloud parameters listed in Table 1.  The corresponding column densities per unit velocity interval are shown in the bottom panel, for $\rm OH^+$ ({\it blue}) and $\rm H_2O^+$({\it green}).  Also shown, for comparison, are the HI ({\it black}; from Fish et al.\ 2003) and para-H$_2$O ({\it red}; Sonnentrucker et al.\ 2010) column densities.  {Where the HI column density exceeds the maximum value on the vertical scale, that value should be regarded as a lower limit, the 21~cm transmission being zero to within the errors.}}

\end{figure}

\section{Discussion}

For diffuse molecular clouds, theoretical models (e.g. van~Dishoeck \& Black 1986) have elucidated the ion-neutral chemistry that leads to the production of oxygen-bearing molecules.  The reaction network is initiated by the cosmic-ray ionization of atomic hydrogen to form H$^+$, which undergoes charge transfer with atomic oxygen in a reaction that is slightly endothermic by the equivalent of $\sim$~230~K.  The resultant O$^+$ ions can then undergo a series of exothermic hydrogen atom abstraction reactions, leading to the OH$^+$, H$_2$O$^+$ and H$_3$O$^+$.  The latter is removed by dissociative recombination, which can be the dominant source of the neutral molecules OH and H$_2$O.  If the ratio of electron density to H$_2$ is sufficiently large, as it is in diffuse clouds of small molecular fraction, the pipeline leading from O$^+$ to OH$^+$ to H$_2$O$^+$ to 
$\rm H_3O^+$ can be a leaky one, with the flow of ionization reduced at each step by the dissociative recombination of OH$^+$ and $\rm H_2O^+$.  In dense molecular clouds, by contrast, the atomic hydrogen abundance and temperature are both too small for O$^+$ production to be efficient, and the dominant source of OH$^+$ is the reaction of H$_3^+$ with O.  Once again, the chemistry is driven by cosmic ray ionization -- the original source of H$_3^+$ -- but now the conversion of OH$^+$ to $\rm H_2O^+$ to $\rm H_3O^+$ proceeds with almost 100$\%$ efficiency. 

As discussed by Gerin et al.\ (2010), the OH$^+$/$\rm H_2O^+$ abundance ratio provides a critical probe of the molecular fraction.  Regardless of the production mechanism for OH$^+$, the OH$^+$/$\rm H_2O^+$ ratio is given by

$${k({\rm H_2 \vert OH^+}) \over k ({\rm H_2 \vert H_2O^+})} + {n(e) k({\rm e \vert H^+}) \over n({\rm H}_2) k({\rm H_2 \vert H_2O^+})} 
=0.64 + 1490\,{x_e T_2^{-0.5} \over f({\rm H}_2)}, \eqno(1)$$
where $k({\rm X \vert Y})$ is the rate coefficient for reaction of X with Y, $n({\rm X})$ is the density of species X, $x_e=n({\rm e})/[2n({\rm H}_2)+ n({\rm H})]$ is the fractional ionization, $f({\rm H}_2) = 2n({\rm H}_2)/[2n({\rm H}_2)+ n({\rm H})]$ is the molecular fraction, and $T = 100\,T_2$~K is the temperature.  Here, we adopt the reaction rate coefficients tabulated in Gerin et al.\ (2010; their Table 3).

This simple expression, which accounts exactly for all results that we have obtained to date with the Meudon PDR model (Le Petit et al. 2006; Goicoechea \& Le Bourlot 2007), assumes only that H$_2$O$^+$ is produced by reaction of OH$^+$ with H$_2$, is destroyed by dissociative recombination or reaction with H$_2$, and has reached a steady-state abundance.  For the six absorption components in which the OH$^+$ and $\rm H_2O^+$ column densities are well determined, the resultant abundance ratios range from $\sim 3 - 15 $,  requiring $f({\rm H}_2)$ in the range $\sim 
0.02 - 0.08$.  Since the largest plausible fractional ionization in a neutral gas cloud is $\sim 1.4 \times 10^{-4},$ corresponding to the complete ionization of carbon, the OH$^+$ ions must reside primarily within clouds of low molecular fraction.  The derived molecular fraction is given in Table 2 for each absorption component.

Our conclusion that OH$^+$ resides primarily in clouds that are predominantly atomic  -- a result obtained previously by Gerin et al.\ (2010) for the sight-line to G10.6--0.4 -- is strongly corroborated by the observed distribution of absorbing material in velocity space (Fig.\ 2, bottom panel).  The distribution of H$_2$O$^+$ (in green) and OH$^+$ (blue) is similar to that of atomic hydrogen, as determined by 21 cm observations reported by Fish et al.\ (2003; black), {(although atomic hydrogen is evidently present without accompanying OH$^+$ and H$_2$O$^+$ at LSR velocities in the range --10 to 0 km~s$^{-1}$).  The distribution of OH$^+$ and H$_2$O$^+$ is} strikingly {\it dissimilar} from that of para-H$_2$O (red), which has been observed by HIFI and reported by Sonnentrucker et al.\ (2010).  Whereas water and other molecules detected previously toward W49N appear to reside primarily in clouds of relatively narrow velocity width, $\rm OH^+$, $\rm H_2O^+$ and H are more broadly distributed in velocity space.

\begin{table}
\caption{$\rm OH^+$ and $\rm H_2O^+$ abundances}
\begin{tabular}{c c c c c c }
\hline\hline
$v_{\rm LSR}$ & $N({\rm H})$ $^a$ & ${\rm OH^+}$/H & ${\rm H_2O^+}$/H & $\zeta_{\rm H}\epsilon / n_2$ \\
$\rm km\,s^{-1}$ & $10^{21}\rm cm^{-2}$ & & & $10^{-16}\rm s^{-1}$ \\
\hline
30 -- 50  &  6.95   &  $5.1 \times 10^{-8}$ & $ 5.1 \times 10^{-9}$ & 1.15 \\
50 -- 78  &  7.23   &  $3.0 \times 10^{-8}$ & $ 2.6 \times 10^{-9}$ & 0.63 \\
\hline\hline
\end{tabular}

$^a$from Godard et al.\ (2010) for an assumed spin temperature of 100~K.
\end{table}

In Table 2, we present the abundances of $\rm OH^+$ and $\rm H_2O^+$ relative to atomic hydrogen in two velocity ranges considered previously by Godard et al.\ (2010).  Here we adopt the estimates of $N({\rm H})$ derived by Godard et al.\ from the HI 21 cm spectra of Fish et al.\ (2003) {for an assumed spin temperature, $T_S$, of 100~K.}  As discussed by Gerin et al.\ (2010), the OH$^+$ to atomic hydrogen ratio probes the cosmic ray ionization rate, $\zeta_{\rm H}$, defined here as the total rate of ionization per hydrogen atom.  The presence of OH$^+$ in clouds of small molecular fraction was anticipated by Liszt (2007), in his investigation of the time-scale for H$_2$ formation in diffuse atomic clouds.   In his models, the hydrogen molecular fraction approaches steady state only after $10^7$ yr. His predicted formation rate of OH$^+$ at $\zeta_H=3\times 10^{-16}$ s$^{-1}$ is sufficient to attain $n({\rm OH}^+)/n_{H}=5\times 10^{-8}$ in clouds at $n_H\sim 50$ cm$^{-3}$ when the cloud age is in the range 2 to 10 million yr. 

Defining $\epsilon$ as the ratio of the OH$^+$ production rate to the cosmic ray ionization rate, assuming that OH$^+$ is destroyed by reaction with H$_2$ and dissociative recombination at a rate equal to its formation rate, and using eqn.\ (1) to relate $n_e/n({\rm H}_2)$ to the observed $\rm OH^+/H_2O^+$ ratio, we find that

$$ {\zeta_H \epsilon \over  n_2} = 9.1 \times 10^{-17} {x({\rm OH^+})T_2^{-0.5} \over 10^{-7}} {x_e \over 1.4 \, 10^{-4}}
\biggl( 1 + {11.6 \over R - 0.64}\biggr) \, \rm s^{-1}, \eqno(2) $$
where $R = n({\rm OH^+})/n({\rm H_2O^+})$,  $x({\rm OH^+}) = n({\rm OH^+})/n_H$, $n_2 = n_{\rm H}/10^2\rm \, cm^{-3}$ and $n_H = n({\rm H}) + 2n({\rm H}_2)$. 
Values for $\zeta_H \epsilon / n_2$ are tabulated in Table 2 for an assumed temperature of 100 K and a fractional ionization of $1.4 \, 10^{-4}$.  If cosmic ray ionization is entirely responsible for the production of OH$^+$, then $\epsilon$ represents the efficiency with which ionization is transferred from H$^+$ to OH$^+$.  In this case, $\epsilon$ must be $\le 1$, and our analysis yields a robust lower limit on $\zeta_{H}$.  

Based upon parameter studies performed with the Meudon PDR model, to be described in detail in a future publication, we find that values of $\epsilon$ close to unity can be achieved under a wide range of conditions typical of diffuse molecular clouds.  We have computed the structure of diffuse molecular clouds for a set of models with {gas-phase carbon and oxygen abundances of $1.38 \times 10^{-4}$ and
$3.02 \times 10^{-4}$ respectively with respect to H nuclei}, 
cosmic ray ionization rates $\zeta_{\rm H}$ of $10^{-17}$, $10^{-16}$, or $10^{-15}\, \rm s^{-1}$, density $n_{\rm H}$ of $10^2$ or $10^3\,\rm cm^{-3}$, and UV field $\chi_{UV}$ of 1, 10, or 100 (normalized relative to the mean interstellar value).  Under most of the conditions that we considered, $\epsilon$ lies in the range 0.5 -- 1.0 whenever $R$ lies in the observed range of 3 -- 15, the only exceptions being cases with (1) $\chi_{UV}=100$ and $\zeta_H = 10^{-17}\, \rm s^{-1}$, for which $\epsilon$ can reach values as large as 3; and (2) $\chi_{UV}=1$ and $n_{\rm H} = 10^3\,\rm cm^{-3}$, for which $\epsilon$ lies in the range $0.02 - 0.1$.  In the first of these exceptional cases, the temperature reaches several hundred Kelvin and the production of OH$^+$ is enhanced by reaction of O with H$_2$ to form OH, followed by photoionization of OH to produce OH$^+$; in the second case, the temperature is too low to permit efficient charge exchange between O and H$^+$.

Given the temperature, $100$~K, and density, $100 \rm \, cm^{-3}$, typical of diffuse clouds with a small molecular fraction, our results suggest that the OH$^+$ and $\rm H_2O^+$ abundances observed toward W49 imply a cosmic ray ionization rate $\zeta_{H} \sim 0.6 - 2.4 \times 10^{-16}\,\rm s^{-1}$ for atomic hydrogen.  This is the total rate of ionization per hydrogen atom, including secondary ionizations.  {The range given here reflects uncertainties in $\epsilon$, assumed to lie between 0.5 and 1, and  variations in $R$ from one component to another; for temperatures and densities different from those assumed above, the derived value of $\zeta_{H}$ would scale as $n T^{-0.5} T_S^{-1}$.}  
This estimate is in good agreement with values for the primary ionization rate per H nucleon obtained by Indriolo et al.\ (2007) from an analysis of H$_3^+$ abundances observed along many sight-lines intersecting diffuse clouds in the Galactic disk; these were in the range $0.5 - 3.2 \times 10^{-16}\,\rm s^{-1}$ whenever H$_3^+$ was detected (although smaller values were allowed but not required in cases where H$_3^+$ was not detected).  A similar estimate ($\zeta_{H} \sim {1.2} \times 10^{-16}\,\rm s^{-1}$) was inferred  by Le Petit, Roueff \& Herbst (2004) from observations of H$_3^+$ and other species toward $\zeta$ Persei.

\appendix

\section{Acknowledgements}

HIFI has been designed and built by a consortium of institutes and university departments from across Europe, Canada and the United States under the leadership of SRON Netherlands Institute for Space Research, Groningen, The Netherlands and with major contributions from Germany, France and the US. Consortium members are: Canada: CSA, U.~Waterloo; France: CESR, LAB, LERMA, IRAM; Germany: KOSMA, MPIfR, MPS; Ireland, NUI Maynooth; Italy: ASI, IFSI-INAF, Osservatorio Astrofisico di Arcetri-INAF; Netherlands: SRON, TUD; Poland: CAMK, CBK; Spain: Observatorio Astron\'omico Nacional (IGN),Centro de Astrobiolog\'a (CSIC-INTA). Sweden: Chalmers University of Technology - MC2, RSS \& GARD;Onsala Space Observatory; Swedish National Space Board, Stockholm University - Stockholm Observatory;Switzerland: ETH Zurich, FHNW; USA: Caltech, JPL, NHSC.

This research was performed, in part, through a JPL contract funded by the National Aeronautics and Space Administration. JRG was supported by a Ram\'on y Cajal contract and by the MICINN/AYA2009-07304 grant.  R.Sz. and M.Sch. acknowledge support from grant N 203 393334 from Polish MNiSW.

\section{$\bf H_2O^+$ spectroscopy}

The rest frequencies for transitions of $\rm H_2O^+$ have been the subject of some controversy (Ossenkopf et al.\ 2010).  The frequency given by M\"urtz et al.\ (1998) for the strongest hyperfine component is $1115.204 \pm 0.002$~GHz. This value, which is based upon laser magnetic resonance (LMR) experiments, is $\sim $ 28 MHz higher than that  predicted by Ossenkopf et al.\ (2010) from an analysis of earlier LMR data (Strahan et al.\ 1986), 
and 42 MHz higher than the "astronomically-determined" rest frequency presented by Ossenkopf et al.\
We have investigated how well the OH$^+$ and H$_2$O$^+$ spectra obtained toward W49N can be reconciled for various assumed values of the $\rm H_2O^+$ transition frequency.  We allowed the latter to vary, in performing the spectral fitting procedure described in \S3 above, but without changing the assumed spacing of the hyperfine components.  Our best fit frequency for the strongest H$_2$O$^+$ hyperfine component was $1115.209$ GHz, a value lying close to that given by M\"urtz et al.\ (1998).  This result would be in error if the OH$^+$ molecules have systematic velocity shift relative to H$_2$O$^+$ within a given cloud.  Also, insofar as the frequency scale is tied to the assumed OH$^+$ rest frequency, the fractional error in the derived H$_2$O$^+$ frequency can be no smaller than that for the OH$^+$ rest frequency.  Thus, our ``astrophysical determination" of the H$_2$O$^+$ rest frequency is entirely consistent with the laboratory value reported by M\"urtz et al.\ (1998).  However, it is apparently inconsistent with the significantly smaller values given by Ossenkopf et al.\ (2010).  In Figure B.1, we show the  reduced $\chi^2$ for the best fit that can be obtained for a given assumed H$_2$O$^+$ rest frequency, as a function of that frequency.  The best fit shows a minimum close to the M\"urtz et al.\ (1998) rest frequency and is significantly worse for the other values.

\begin{figure}
\includegraphics[width=9 cm]{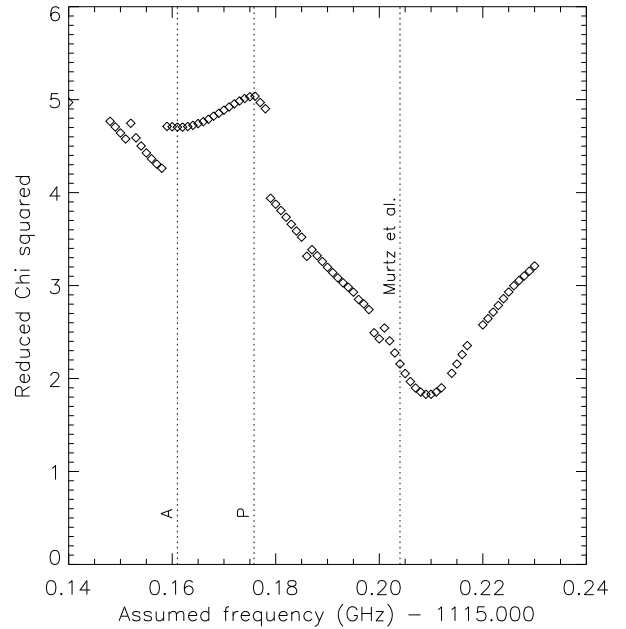}
\caption{Reduced $\chi^2$ for the best fit, as a function of the frequency assumed for the strongest H$_2$O$^+$ hyperfine component.  Dotted lines indicate the values given by M\"urtz et al.\ 1998 and Ossenkopf et al.\ (2010) (P=predicted value based upon Strahan et al.\ 1986 spectroscopy; and A=astronomically-determined value)}
\end{figure}

\end{document}